\documentclass{KapProc} 

\usepackage{epsf}
\usepackage{graphicx}
\normallatexbib

\begin{document}

\articletitle{Zero-bias transport anomaly in metallic nanobridges}

\articlesubtitle{Magnetic field dependence and universal conductance
fluctuations}

\author{H. B. Weber$^a$, R. H\"aussler$^b$,  H. v. L\"ohneysen$^{a,b}$
and  J. Kroha$^c$}
\affil{$^a$  Forschungszentrum Karlsruhe, Institut f\"ur
Nanotechnologie, D-76021 Karlsruhe\\
$^b$ Physikalisches Institut, Universit\"at Karlsruhe, D-76128 
Karlsruhe, Germany\\
$^c$ Institut f\"ur Theorie der Kondensierten Materie, 
Universit\"at Karlsruhe\\ \hspace*{0.3cm}D-76128 Karlsruhe, Germany}

\email{Heiko.Weber@int.fzk.de}

\chaptitlerunninghead{Zero-bias transport anomaly in metallic nanobridges}

\begin{keywords}
Electron-electron interaction, diffusive transport, non-equilibrium
\end{keywords}

\begin{abstract}
We present data of transport measurements through a metallic nano\-bridge 
exhibiting diffusive electron transport. A logarithmic temperature dependence 
and a zero-bias anomaly in the differential conductance are observed,
independent of magnetic field. 
The data can be described by a single scaling law. 
The theory of electron-electron interaction in disordered systems,
adapted to the case of finite-size systems in non-equilibrium,
yields quantitative agreement with experiment. Measurements of universal
conductance functuations support the assumptions of the theory
about the electronic phase coherence.
\end{abstract}

It is well known that in bulk metals and semiconductors with
diffusive transport the electron-electron interaction causes 
an anomaly in the electronic density of
states (DOS) at the Fermi level.  
As explained by Aronov and Al'tshuler (A-A) in the
1980s \cite{AA.85,LeeRama}, this correction  is  induced by the long-range, retarded
character of the dynamically screened Coulomb interaction in a diffusive
system. It has been observed in thermodynamic equilibrium
by tunneling spectroscopy on disordered metals \cite{Imry,hertel}.
In the present article we address the question
how this anomaly is modified in a nanoscopic sample or metal 
bridge whose size $L$ is smaller than the dephasing length $L_{\varphi}$ 
(and all inelastic relaxation lengths), in particular when it
is driven out of equilibrium by a finite 
bias voltage $U$ applied between the ends of the bridge.
\begin{figure}[ht]
\centerline{
\epsfysize=3.8cm
\epsfbox{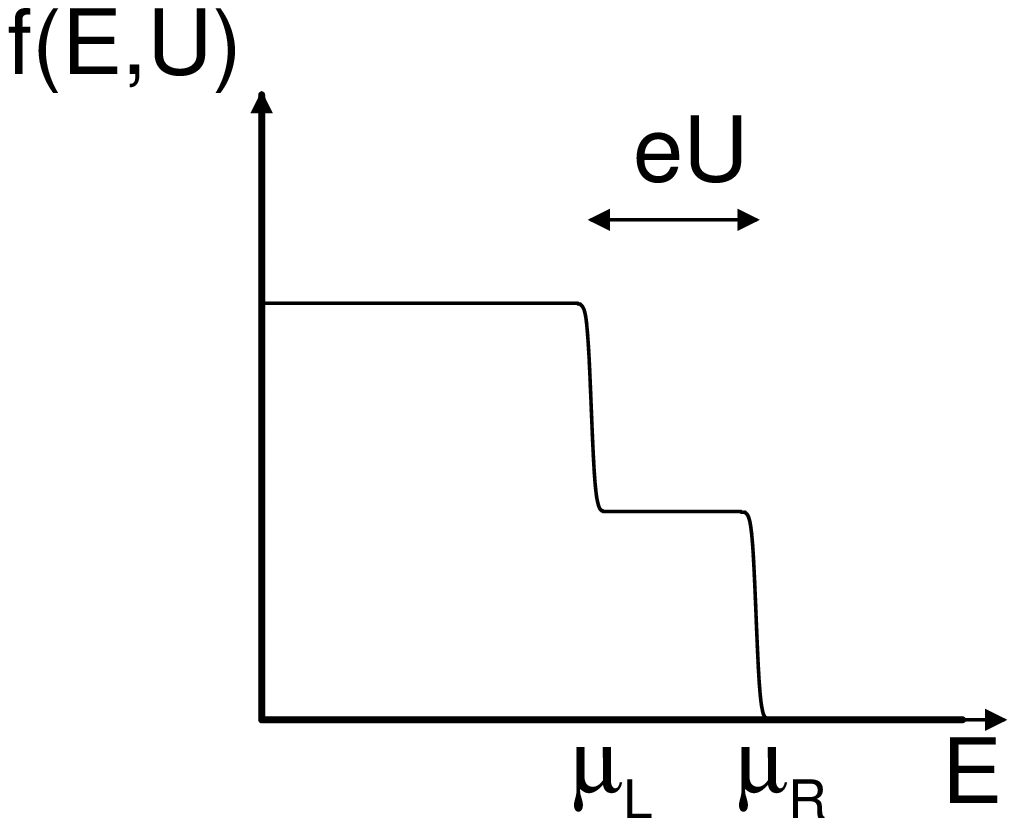}
\hfill
\epsfxsize=5.8cm
\epsfbox{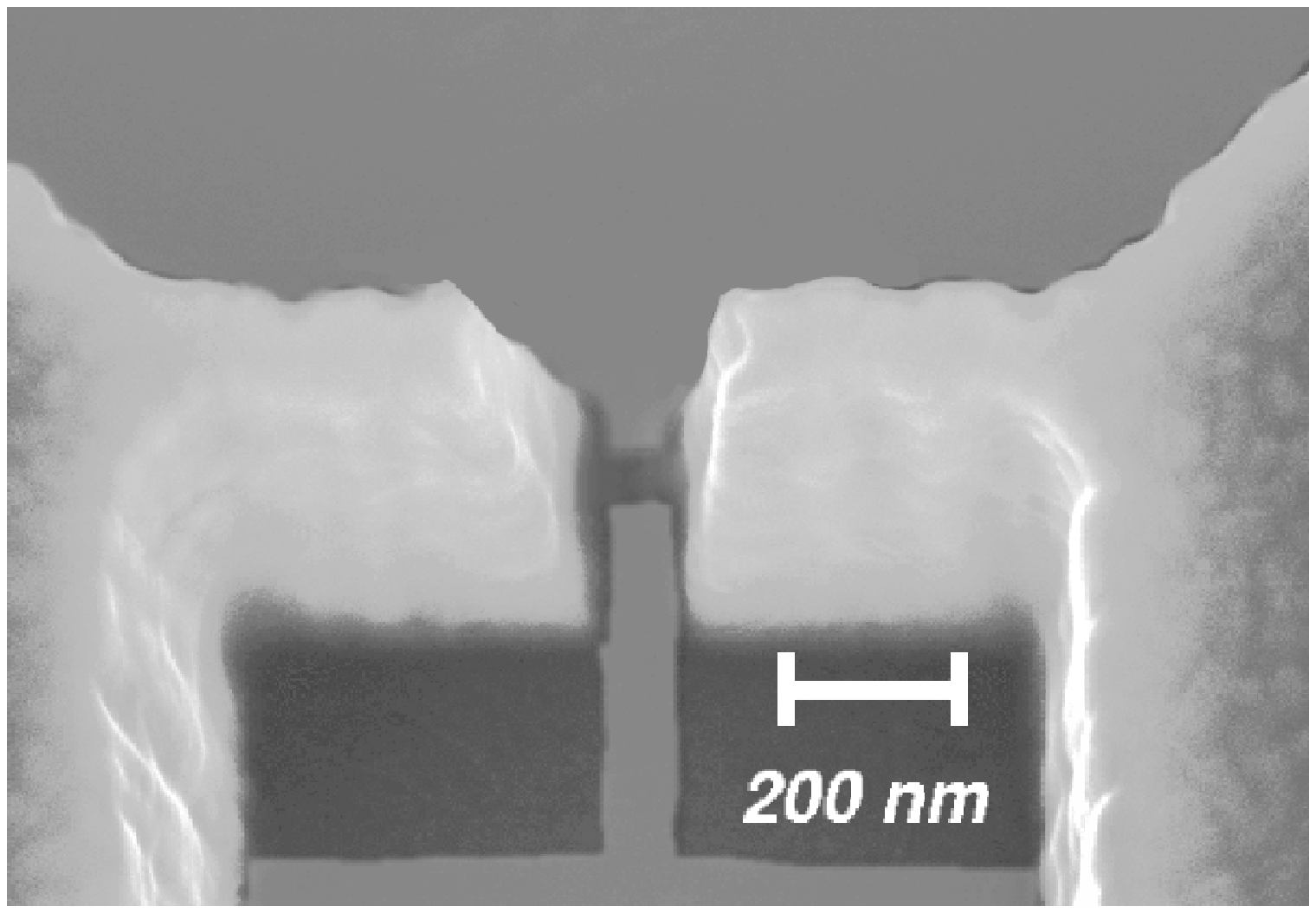}
}
\dblcaption{\label{fig:double} 
Quasiparticle distribution function $f(E,U)$
in a nanoscopic sample ($L < L_{\varphi}$)
with finite bias $U$ applied. The spacing between 
the steps is $eU$.
}
           {\label{fig:weber_sample} 
SEM picture of a typical sample. The bridge (10 nm thick) appears dark,
the reservoirs are 70 times thicker.
The  thick Cu replica of the bridge, 
which is not connected to the electrodes, is removed from the
picture for clarity.}
           \end{figure}
\inxx{captions,figure}     
Since in this situation
energy relaxation is negligible for electrons traversing the bridge,
no local equilibrium is reached at any point in the bridge.  
Rather, the electron liquids penetrating from the left and right 
leads into the bridge
remain at their respective electrochemical potentials $\mu_L$ and $\mu_R$. 
Consequently, the
quasiparticle distribution function is a linear superposition of Fermi
distributions in the left and right leads and displays a double-step form 
(see Fig. \ref{fig:double}). 
This non-equilibrium distribution 
has been suggested theoretically \cite{kulik} and 
has recently been observed experimentally by tunneling spectroscopy 
\cite{pothier.97} (where in
addition the steps were rounded due to interactions in long wires).
It should be distinguished from the hot electron regime \cite{nagaev.95}, 
where local thermalization in a current-carrying system occurs.

In this paper we report on the observation 
of a zero-bias conductance anomaly (ZBA) in
metallic nanobridges \cite{aafilm} which, by their special design, 
allow to establish the well-definied non-equlibrium described above.
The ZBA is characterized by a logarithmic
scaling law, independent of an applied magnetic field. 
We show that the ZBA, including the scaling behavior, 
can be explained in detail \cite{aafilm} 
via a Landauer-B\"uttiker formula \cite{wingreen.92}
as arising from
the A-A correction to the electronic DOS of the bridge in non-equilibrium. 
The independence of the data of magnetic field allows us to distinguish the A-A
anomaly from various other effects, like weak localization (WL) \cite{Schmid}
and magnetic impurities, which might cause a ZBA as well.
We also present measurements of
universal conductance fluctuations (UCF) in one of the 
nanobridges. They confirm that 
the phase coherence extends over the entire nanobridge, which is
the criterion for the Landauer-B\"uttiker approach to be applicable.

The data shown in the following are obtained from one sample of
a Cu$_{82}$Au$_{18}$ nanobridge.
We have investigated other Cu$_{100-x}$Au$_{x}$ and Cu bridges as well, 
with very similar results. Experimental details of the fabrication and the
measurement are described in Ref.~\cite{aafilm}.
The bridge is $L=80$~nm long, about 80~nm wide and has a thickness 
of $d=10$~nm. It is placed in 
good metallic contact between two bulk Cu leads, which are about 70 times
thicker than the bridge (see Fig.\,\ref{fig:weber_sample}) and extend 
over a large area
of about 1~mm$^2$ each. 
Hence the voltage applied to the sample drops only along the bridge, 
and the Joule heating power is reliably conducted away by the leads.
The mean free path in the sample is about $\ell=6.5$~nm,
corresponding to a diffusion constant of $D = 34{\rm cm^2/s}$, i.e. it is 
comparable to the thickness $d$, but much shorter than the lateral length $L$.
Therefore, the electronic density modes in the bridge obey the rules of  
two-dimensional (2D) diffusive motion. 
The 2D design also allows to distinguish the A-A conductance anomaly
from a possible two-channel Kondo (TCK) 
effect induced by two-level systems \cite{zawa.98}, 
which has been put forward as the origin of ZBAs
observed in ultrasmall point contacts \cite{ralph.94}: 
In 3D both the A-A and the TCK anomalies   
show square-root power-law behavior; in 2D the A-A correction 
is logarithmic, while the TCK singularity, as a local effect, 
is independent of dimension.

\begin{figure} 
\epsfxsize7.7cm \centering\leavevmode\epsfbox{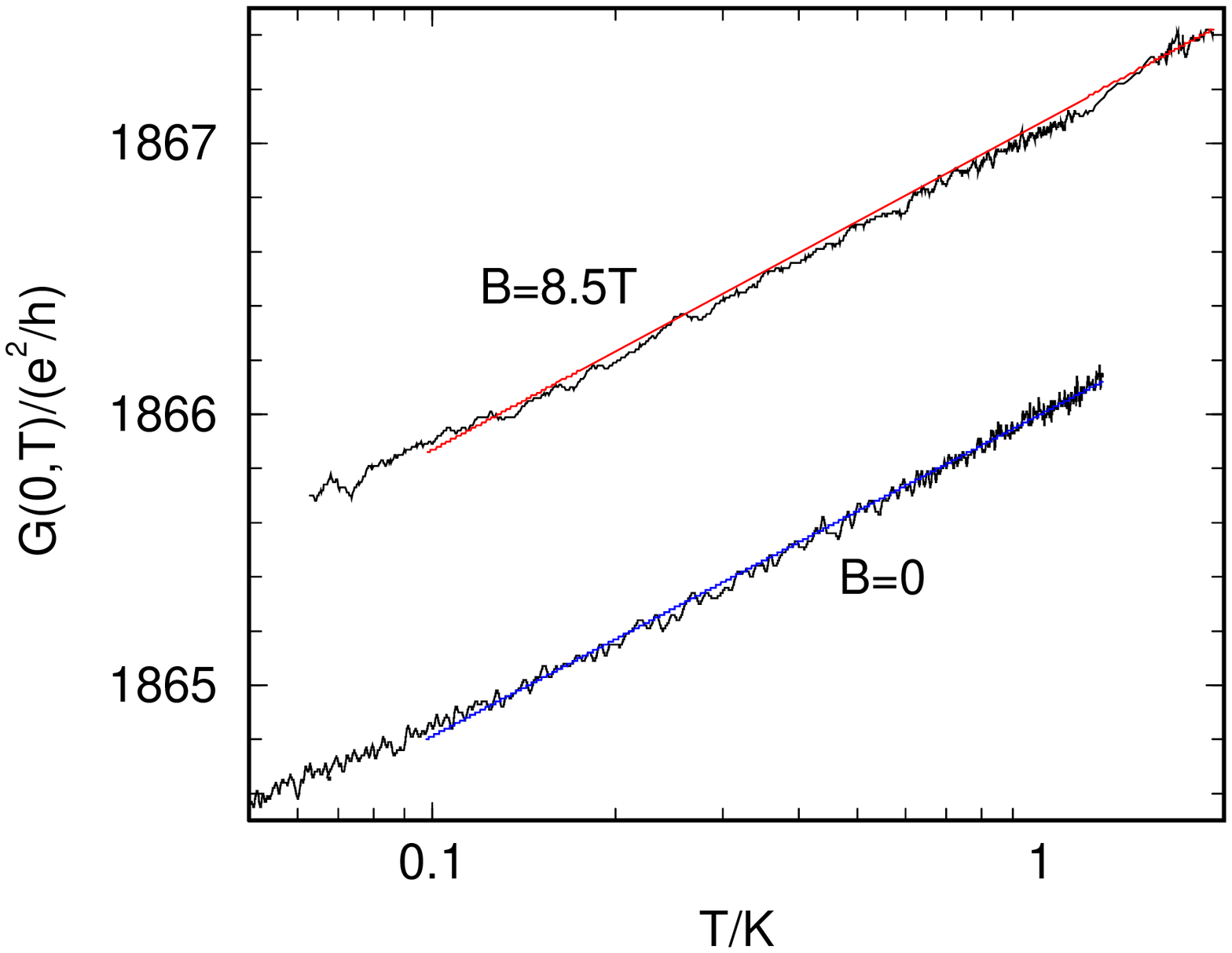}  
\vspace*{-0.3cm} 	
\caption{Zero-bias conductance at $B=0$ and $B=8.5$~T. 
The magnetic field is applied perpendicular to the film. 
The constant background value of the conductance 
changes due to universal conductance fluctuations.} 	
\label{weber_conductance.8.5T} 
\end{figure}
\begin{figure} 
\epsfxsize7.cm \centering\leavevmode\epsfbox{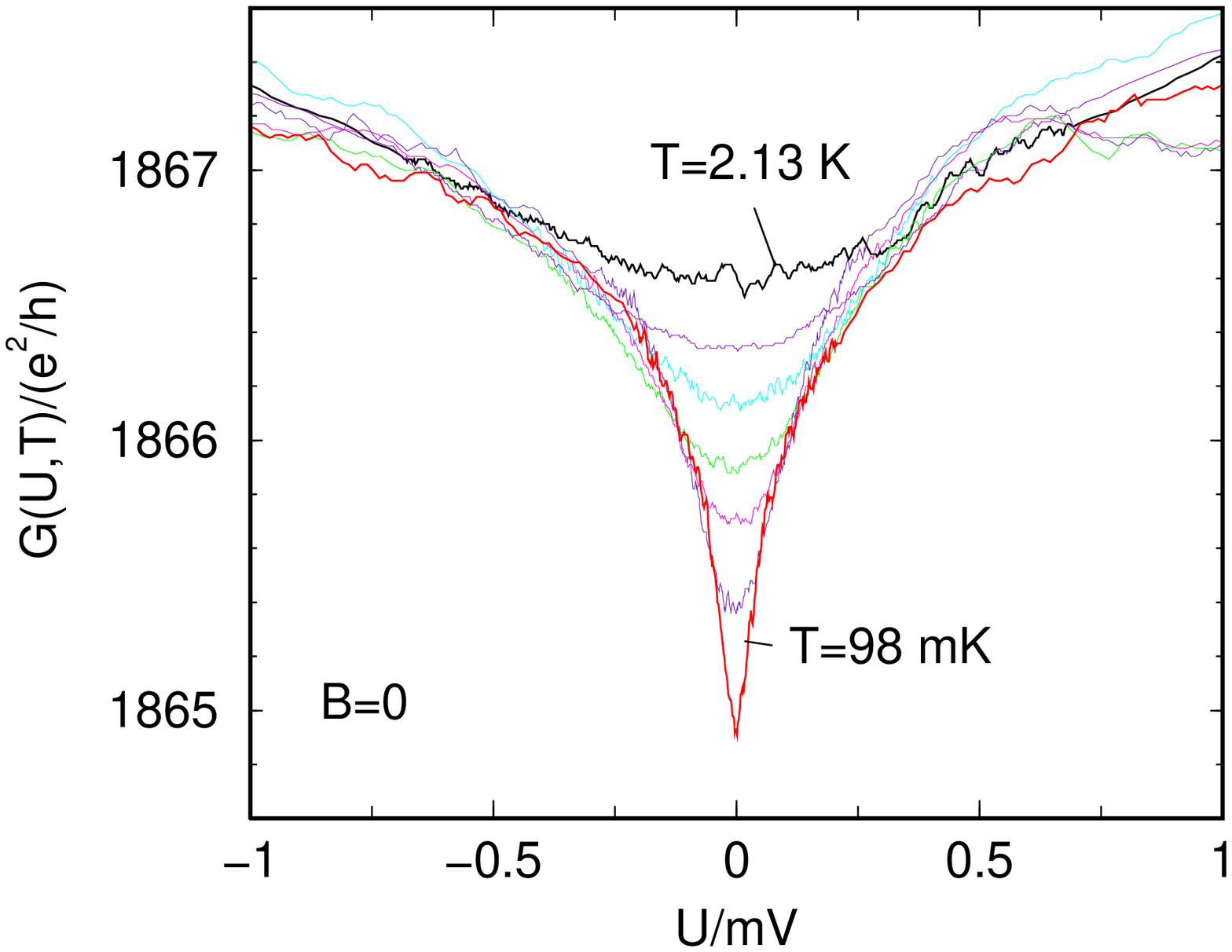}  
\vspace*{-0.3cm} 	
\caption{Raw conductance data of the Cu$_{82}$Au$_{18}$ bridge
for various fixed temperaures, $T$=0.098~K, 
0.25~K, 0.5~K, 0.75~K, 1.0~K, 1.5~K, 2.13~K.} 	
\label{weber_voltagedependence} 
\end{figure}

\begin{figure} 
\vspace*{-2.0cm}
\epsfxsize11.0cm 
\centering\leavevmode\epsfbox{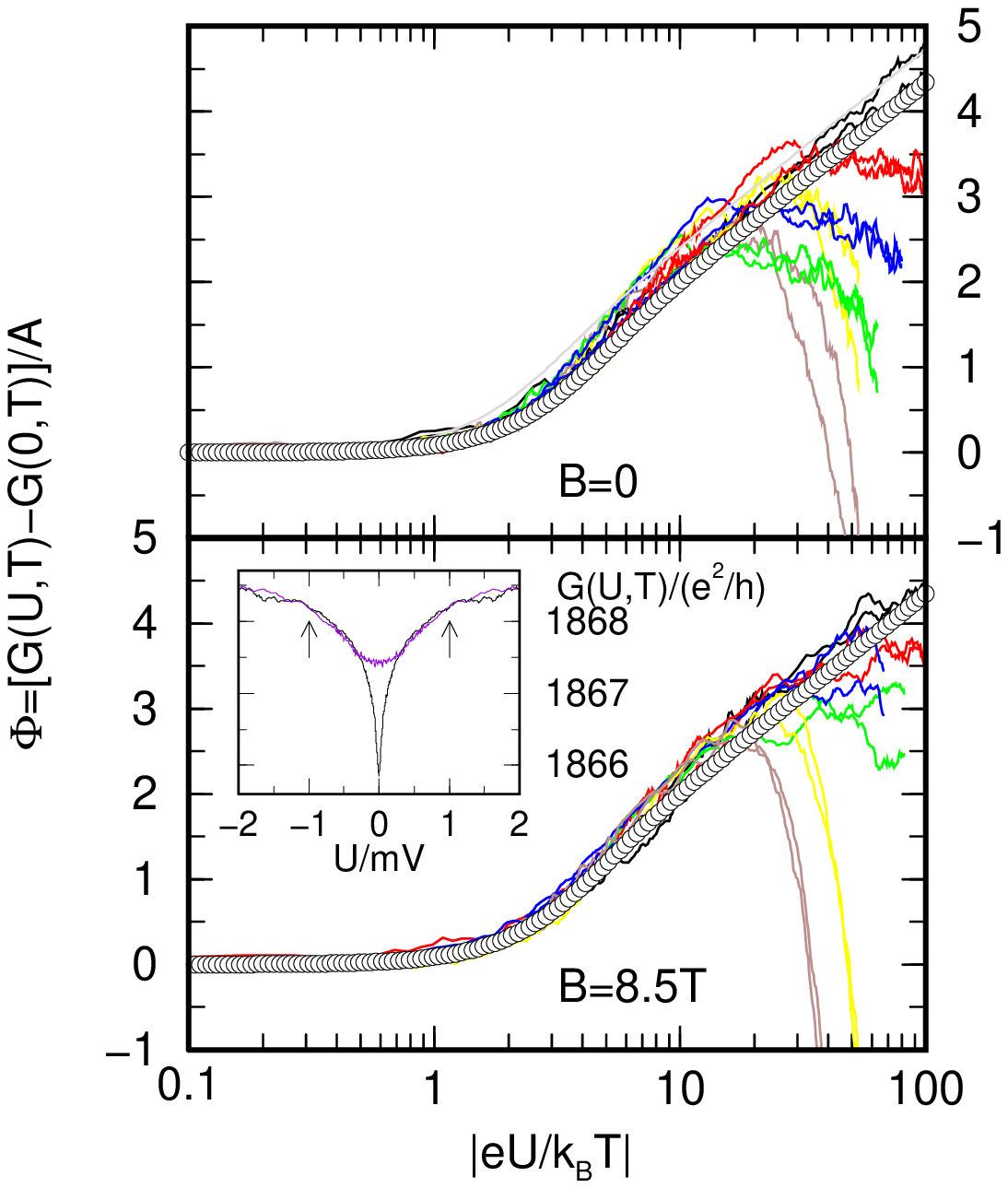}  
\vspace*{-0.5cm} 	
\caption{Scaling plot of the conductance at magnetic field $B=0$ (top)
and at $B=8.5$~T (bottom). 
Lines represent experimental data for various fixed 
temperatures as in Fig.~\ref{weber_voltagedependence}, where
measurements both at positive and at negative bias are included.
In both plots, circles represent theoretical calculations of the leading
A-A conductance correction following from Eq.~(\ref{eq:deltaNsc}), 
which is independent of magnetic field $B$. 
The inset shows raw data of the ZBA in a magnetic field of $B=8.5$~T
for temperatures $T=100$~mK and $T=2$~K, where
the position of the Zeeman energy (see text) is marked by arrows.
The $B$ independence of the experimental data is clearly seen.} 	
\label{weber_scaling} 
\end{figure}

 
\noindent
We observe a logarithmic temperature $T$ 
dependence of the zero-bias conductance 
$G(0,T)=G_0+A\cdot {\rm ln}(T/1{\rm K})$ in a range of $T= 100$~mK to 2.1~K, 
with an amplitude of $A = 0.49~e^2/h$, as shown in  
Fig.~\ref{weber_conductance.8.5T} for vanishing magnetic field, $B=0$
(lower curve). 
Below 100~mK, the data deviate somewhat from this 
logarithmic behavior, a fact that we attribute to incomplete thermalization.  
It is seen that the amplitude $A$ is independent of magnetic field. 
When applying a finite bias voltage $U$, a small, voltage-symmetric anomaly 
in the conductance  was found in the differential conductance 
$G(U,T)$ at low bias. Experimental raw data are shown 
in Fig. \ref{weber_voltagedependence}. 
When the temperature is lowered, the anomaly gets more pronounced. 
It can be characterized by a striking scaling property: When the zero-bias 
conductance $G(0,T)$ at the respective temperature is subtracted 
from $G(U,T)$, the data displayed as a function of $eU/k_BT$, 
collapse onto one single scaling curve 
in a wide region around zero bias.
Moreover, after normalizing the conductance with the amplitude $A$ 
of the $T$ dependent linear response signal (see above), 
$(G(U,T)-G(0,T))/A$ is nearly identical for all investigated samples, 
where the mean free path $\ell$ and the thickness $d$ 
were varied within a factor of two.
Hence, all the conductance data $G(U,T)$ can be described by a single 
scaling law,
\begin{equation}
 G(U,T)=G_0+A\cdot {\rm ln}(T/1{\rm K})+A\cdot \Phi (eU/k_BT)\, .
\end{equation}
The scaling function $\Phi(x)$ obtained in this way is displayed 
in Fig. \ref{weber_scaling}, where the asymptotic behavior 
$\Phi (x) = {\rm ln}x$ for $x\gg 1$ may be extracted. 
  \noindent

When a perpendicular magnetic field $B=8.5$~T is applied,  
the scaling behavior persists, with the amplitude $A$ 
(Fig.~\ref{weber_conductance.8.5T}) and, moreover,
the scaling function $\Phi (x)$ (Fig.~\ref{weber_scaling}) remaining unchanged. 
This is clear evidence that WL is not observed in our measurements,
as might already have been expected from the shortness of our samples. 
 \noindent
Logarithmic behavior may also be caused by magnetic impurities or by
non-magnetic TCK defects \cite{zawa.98} above their respective Kondo
temperatures $T_K$. Since an applied field of $B=8.5$~T does not modify
the ZBA, any magnetic impurities present in the sample must 
have $T_K \gg 8.5$~K. However,
the logarithmic behavior of the zero-bias conductance observed down to the 
lowest $T$ (Fig.~\ref{weber_conductance.8.5T}) puts an upper bound to the Kondo 
temperature, $T_K<0.1$~K, thus ruling out magnetic impurities 
as the origin of the ZBA.
In the TCK scenario, from point-contact spectroscopy on Cu one expects  
$T_K \approx$ 5 to 10~K \cite{ralph.94,hettler.94}. Hence, it is unlikely
that the ZBA is due to TCK defects for the same reason as in the magnetic
case. The assumption that there is no sizable number of TCK defects 
present in our Cu$_{82}$Au$_{18}$ samples is consistent with the
fact that in Cu point contacts investigated previously the 
TCK signal completely disappeared upon doping with 1\%  Au or more
\cite{ralph.94}.
Because of the good metallic contact between bridge and leads, 
charging effects at the 
interfaces \cite{nazarov.92} may be regarded as negligible in our devices.

\begin{figure} 
\epsfxsize8.5cm \centering\leavevmode\epsfbox{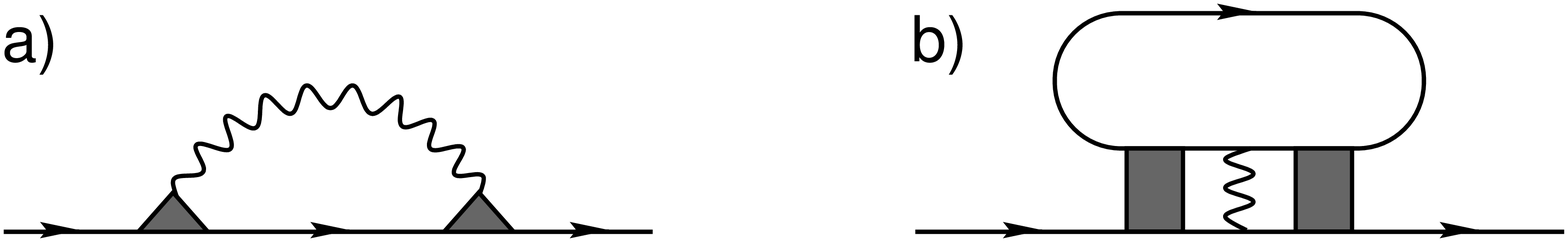}  
\caption{a) Exchange and b) Hartree diagram describing the A-A DOS
correction to leading order in the effective Coulomb interaction
$\bar v_q$ (wiggly line).
Shaded triangles and squares represent the diffusion density and 
particle-hole vertex, respectively.} 	
\label{fig:diagrams} 
\end{figure}

In order to understand the logarithmic ZBA theoretically, it is
important to note that the length $L$ of our
disordered nanobridges is small compared to the
dephasing length $L_{\varphi} \sim \sqrt{\hbar D /k_BT}$ 
and all inelastic relaxation lengths, as will be verified below. 
Hence, the electrons occupy the exact
single-particle eigenstates of the disordered bridge while traversing
the system, i.e.~the DC transport is ballistic (i.e.~zero-dimensional),
since it involves only zero-frequency modes,
even when a finite bias voltage is applied. 
In this situation the Landauer-B\"uttiker approach is applicable,
where the conductance is expressed in terms of the exact eigenstates
or channels of the transmitting region, and which has been generalized
to interacting systems by Meir and Wingreen \cite{wingreen.92}.
The current through the bridge at bias $U$ thus reads,
\begin{equation}
I(U)=\frac{e}{\hbar} 
\Gamma \int \Bigl[ f^0 \Bigl( E-\frac{eU}{2}\Bigr)-f ^0 
\Bigl( E+\frac{eU}{2}\Bigr)\Bigr]\; N(E,U)\; dE,
\label{eq:current}
\end{equation}
where $f^o(E ) = 1/({\rm e}^{E /k_BT}+1)$ 
is the Fermi function and,
for simplicity, the effective lead-to-bridge coupling $\Gamma$ 
is taken to be energy independent and symmetrical for left and right leads
(the more general case is treated in Ref.~\cite{aafilm}).
The energy $E$ is measured with respect to $\mu = (\mu_R +\mu_L)/2$ 
Since the bridge is phase coherent, the quasiparticle distribution function
in the bridge is uniform in space and has the double-step form \cite{kulik}
(Fig.~\ref{fig:double}),
\begin{equation}
f (E ) = \frac{1}{2}\Bigl[ f^0 \Bigl( E-\frac{eU}{2}\Bigr)+f ^0 
\Bigl( E+\frac{eU}{2}\Bigr)\Bigr]\; .
\label{eq:f}
\end{equation}
According to Eq.~(\ref{eq:current}) the current is expressed in terms
of the DOS $N(E,U)$ in the bridge, which in the interacting case may be strongly
affected by the non-equilibrium distribution. In fact, diffusive
density modes exist at finite (2D) wave numbers $q$, $2\pi / L <q<2\pi /\ell$,
and at frequencies $\Omega $ up to the elastic scattering rate 
$1/ \tau = v_F / \ell \approx 0.2 \,{\rm fs} ^{-1}$, although the
DC transport is ballistic. These diffusion
modes couple to the electronic DOS via the dynamically screened Coulomb
interaction and thus give a singular correction to the conductance,
as described by A-A in equilibrium \cite{AA.85}. 
The corresponding DOS corrections are shown diagrammatically to leading
order in the effective electron-electron interaction in Fig.~\ref{fig:diagrams}.
It seen that in the exchange diagram the dynamically screened
Coulomb interaction $\bar v_q(\Omega )$ enters, 
while the Hartree diagram contains the 
statically screened Coulomb interaction $\bar v_q(0)$ because of 
energy conservation at the impurity vertices. 
In an infinite system,
the dynamically screened Coulomb interaction, combined with the
diffusive vertex corrections (shaded triangles in Fig.~\ref{fig:diagrams}a)),
exhbits a hydrodynamic divergence ($q\to 0$, $\Omega \to 0$), 
while the statically
screened one remains finite. Therefore, for a long-range bare interaction
like the Coulomb interaction,
the exchange contribution is always more strongly singular than 
the Hartree term \cite{LeeRama}.

In an infinite 2D film, the exchange term has logarithmic divergences
both in the integral over the frequency $\Omega $ and over the wave number
$q$ transferred by interaction, leading to
 \noindent
the well-known DOS correction  $\delta N(E) 
\propto - {\rm ln}(E /\hbar\tau)\;{\rm ln}(E /\hbar \kappa ^2 D )$, where
$\kappa $ is the inverse 2D screening length \cite{AA.85}. 
In our finite-size 2D bridge the divergence in $q$ is cut
off by the inverse system size both in the Hartree and in the exchange
contribution. It is transformed into a constant 
term ${\rm ln} ({\rm max}(d,\ell)/\ell)$
which stems from the crossover to 3D behavior at short
distances \cite{aafilm}. However, in the exchange term the divergence of
the $\Omega $ integral persists and dominates the Hartree term even in a
finite system. Consequently, near the Fermi step(s), i.e. for energies 
$|E| \stackrel {<}{\sim} (2\pi)^2 E_{Th}$, with $E_{Th}$ the Thouless energy,
simple log behavior instead of log$^2$ behavior remains \cite{aafilm}. 
The corresponding DOS correction may be cast into the scaling form
\begin{eqnarray}
\label{eq:deltaNsc} 
\delta N (y,T) =
\frac{{\rm ln}({\rm max}(d/\ell,1))}{\pi^2E_{Th}} 
\Bigl[ {\rm ln} (T\tau) + \int 
du \Bigl(-\frac{d\bar f(u-y)}{du}\Bigr) 
{\rm ln}|u|  \Bigr] \; , 
\end{eqnarray}
where $\bar f(u)=f(\hbar \Omega /k_B T)$ is the (non-equilibrium)
distribution function in terms of
the dimensionless energy, and $y=E /k_B T$.
Eq.\,(\ref{eq:deltaNsc}) displays two logarithmic
singularities corresponding to the Fermi steps at $y=0$ and $y=-eU/k_BT$. 
It is characteristic for logarithmic behavior that the prefactor of the
term depending on the dimensionless energy $y$ 
is independent of $T$ (in contrast to, e.g.,
power-law scaling) and is equal to the amplitude  of the 
$T$-dependent term at $y=0$. 
Obviously, the universality of $\delta {\cal N}(y,T)$ is preserved
when the differential conductance correction $\Phi (eU/k_BT)=[G(U,T)-G(0,T)]/A$
is calculated using Eqs.~(\ref{eq:current}), 
(\ref{eq:deltaNsc}) and (\ref{eq:f}). The resulting scaling curve is shown
in Fig.~\ref{weber_scaling} and agrees quantitatively with the experimental
data, both without and with applied magnetic field. 
The magnetic field independence of the experimental scaling curves
as well as of the amplitude $A$ (Fig.~\ref{weber_conductance.8.5T}) 
is expected from the A-A anomaly \cite{LeeRama,note}: 
The dominating exchange contribution
(Fig.~\ref{fig:diagrams}a)) is independent of magnetic field since here
diffusion modes enter only through density vertices (shaded triangles), 
which conserve spin. Zeeman splitting 
of the diffusion modes occurs only in the particle-hole vertices  
(shaded squares in Fig.~\ref{fig:diagrams}b)) with opposite
particle and hole spins appearing in the Hartree term, which is negligible
(see above). The position of the
Zeeman splitting energy $\hbar \omega _s = g\mu_B B$ is marked in
the inset of Fig.~\ref{weber_scaling}, where the experimental data
show no structure, as expected. 
\begin{figure} 
\epsfxsize7.0cm \centering\leavevmode\epsfbox{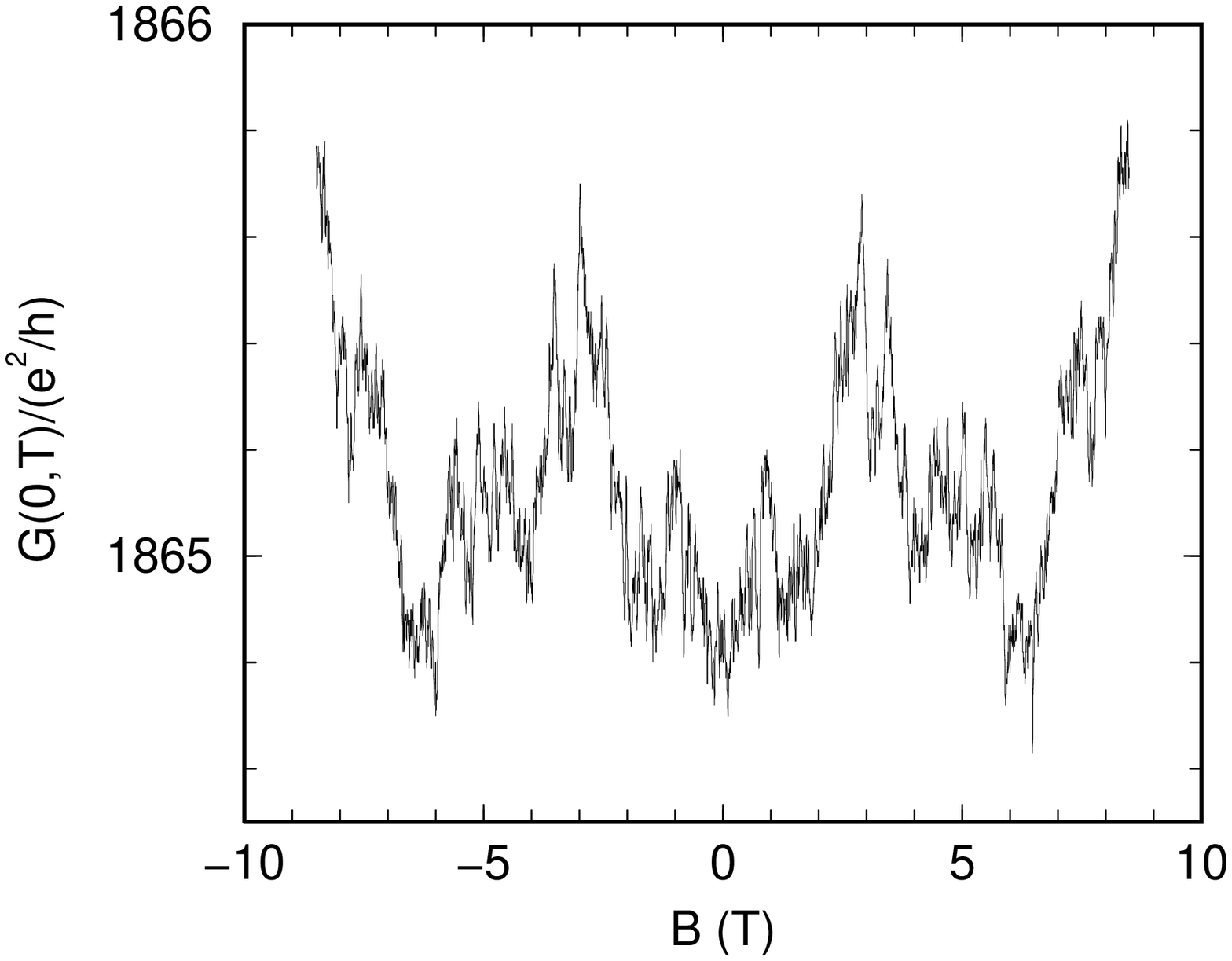}  
\caption{Zero-bias conductance as a function of magnetic field at $T=100$~mK: 
Universal conductance fluctuations.} 	
\label{weber_figucf} 
\end{figure} \noindent

Direct insight in the coherence properties of the samples may be obtained
by investigating the universal conductance fluctuations (UCF).
The magnetoconductance of our sample at $T = 100$~mK is shown 
in Fig. \ref{weber_figucf}. The conductance fluctuations are 
are reproducible and symmetric with respect to reversal of the magnetic 
field, and are, thus, identified as UCF. 
The statistics of the UCF can be analyzed in a standard way
by means of the autocorrelation function
\begin{equation}
C(\Delta B)=\frac{1}{2 B_o}\int_{- B_o}^{B_o}
\delta G(B^\prime) \delta G(B^\prime + \Delta B)dB^\prime
\end{equation}
with $\delta G(B) = G(B)-\langle G \rangle$ and $B_o=8.5$~T. 
The rms amplitude $\delta G_{rms}=\sqrt{C(0)}$ 
is $\delta G \approx 0.22~e^2/h$ for the data shown in Fig.~\ref{weber_figucf}.
From the HWHM of the autocorrelation function 
we obtain a correlation field 
of $B_c = 640$~mT. Another, similar sample of the same size yields 
$\delta G \approx$~0.31 $e^2/h$ and $B_c =1300$~mT. 
The scatter of these data is not surprising: The range of 
experimentally applied fields, $0\leq B \leq 8.5$~T
is not very large compared to the correlation 
field $B_c$, so that no complete averaging over the microscopic 
phase configurations is obtained, and fluctuations in $B_c$ are
expected to be sizable.   
The amplitude of the effect is compatible with other experimental 
data on diffusive metal bridges and theoretical predictions 
\cite{Meir.89}. On the other hand,
the correlation field $B_c$ should be inversely proportional 
to the phase coherent area $A_\phi$: 
$B_c=c \cdot {\Phi_o}/{A_\phi}$ \cite{Lee}, 
$\Phi _o$ being the flux quantum and $c$ 
a constant of ${\cal O}(1)$. Indeed, for our sample, 
a coherence field of $B_c=610$~mT ($B_c=500$~mT for the second sample) 
results when $A_\Phi$ is taken as the bridge area, and
the reservoir-like leads are assumed not
to contribute to the phase coherent area. This agrees  
well with the value of $B_c$ obtained from the UCF 
analysis and, therefore, supports our analysis in terms of  
the Landauer-B\"uttiker approach: The bridge is phase coherent over
its whole spatial extent and the bridge eigenstates are 
well separated from the leads. 

In conclusion, we have shown measurements on a nanoscale, fully 
phase-coherent, metallic nanobridge placed between two reservoir-like leads. 
The particular design allows us to establish a well-defined 
electronic non-equilibrium when a finite bias voltage is applied, 
corresponding to a 
double-step in the electronic distribution function. We observed 
logarithmic $T$ dependence of the zero-bias conduction and logarithmic $U$ 
dependence of the differential conductance, which can be combined into a 
single scaling law. The theory of electron-electron interaction in 
diffusive systems was adapted to the 
constrained bridge geometry, taking the non-equilibrium 
situation fully into account. Instead of a single anomaly in the density 
of states at the Fermi level, well known in equilibrium, two anomalies 
evolve at the two Fermi steps. The theoretical scaling function 
$\Phi (eU/k_BT)$ coincides quantitatively with the experimental data
without adjustable parameter. We also presented the magnetic field 
dependence of the data, showing universal conductance fluctuations, but no 
change in the zero-bias anomaly, in full agreement with the theoretical 
description. 

We are grateful to A.~Mirlin, H.~Pothier, B.~L.~Al'tshuler and
P. W\"olfle for stimulating
discussions. This work was supported by DFG through SFB 195.





%





\begin{chapthebibliography}{99}
\bibitem{note} Magnetic field dependence of the tunneling DOS 
entering Eq.~(\ref{eq:current})
does occur in the presence of strong spin flip scattering or when 
in the case of strong disorder 
the Cooperon correction must be taken into
account in the impurity scattering vertex. Both effects are not expected
to occur in our samples. In the linear-response conductivity,
on the other hand, magnetic field dependence enters
through the Hartree contributions \cite{LeeRama} as observed
in semiconductors \cite{blaschette}, but should be suppressed
in metals due to strong static screening of the Coulomb interaction 
in the latter materials.
\end{chapthebibliography}
\end{document}